\RequirePackage{fix-cm}

\documentclass[smallextended]{svjour3}

\usepackage{graphicx}
\usepackage{color}

\usepackage{float}

\usepackage{amsmath}
\usepackage{amssymb}
\usepackage{bbm}
\usepackage[misc,geometry]{ifsym} 

\usepackage{booktabs}
\usepackage{longtable}

\usepackage{pdflscape}
\usepackage{afterpage}
\usepackage{capt-of}

\usepackage{url}

\usepackage{comment}

\usepackage{cprotect}

\usepackage{listings}

\journalname{Data Mining and Knowledge Discovery}

\begin{document}

\title{\textit{catch22}: CAnonical Time-series CHaracteristics}
\subtitle{\textit{selected through} highly comparative time-series analysis}


\author{Carl H Lubba \and Sarab S Sethi \and Philip Knaute \and Simon R Schultz \and Ben D Fulcher* \and Nick S Jones*
}


\institute{Carl H. Lubba \at
          	Department of Bioengineering, Imperial College London, South Kensington, London SW7 2AZ, UK
			\email{c.lubba15@imperial.ac.uk}
           	\and
           	Sarab S. Sethi \at
          	Department of Mathematics, Imperial College London, South Kensington, London SW7 2AZ, UK
			\email{s.sethi16@imperial.ac.uk}
           	\and
            Philip Knaute \at
          	Department of Mathematics, Imperial College London, South Kensington, London SW7 2AZ, UK
			\email{philip.knaute@gmail.com}
           	\and
            Simon R. Schultz \at
          	Department of Bioengineering, Imperial College London, South Kensington, London SW7 2AZ, UK
			\email{s.schultz@imperial.ac.uk}
           	\and
            Ben D. Fulcher$^\ast$ (\Letter) \at
          	School of Physics, Faculty of Science, Sydney University, Camperdown, NSW 2006, Australia
			\email{ben.fulcher@sydney.edu.au}
           	\and
            Nick S. Jones$^\ast$ (\Letter) \at
          	Department of Mathematics, Imperial College London, South Kensington, London SW7 2AZ, UK
			\email{nick.jones@imperial.ac.uk}
			\and
			$^\ast$: These authors contributed equally to this study.}

\date{Received: date / Accepted: date}

\maketitle

\begin{abstract}
Capturing the dynamical properties of time series concisely as interpretable feature vectors can enable efficient clustering and classification for time-series applications across science and industry.
Selecting an appropriate feature-based representation of time series for a given application can be achieved through systematic comparison across a comprehensive time-series feature library, such as those in the \textit{hctsa} toolbox.
However, this approach is computationally expensive and involves evaluating many similar features, limiting the widespread adoption of feature-based representations of time series for real-world applications.
In this work, we introduce a method to infer small sets of time-series features that (i) exhibit strong classification performance across a given collection of time-series problems, and (ii) are minimally redundant.
Applying our method to a set of 93 time-series classification datasets (containing over 147\,000 time series) 
and using a filtered version of the \textit{hctsa} feature library (4791 features), we introduce a generically useful set of 22 CAnonical Time-series CHaracteristics, \emph{catch22}.
This dimensionality reduction, from 4791 to 22, is associated with an approximately 1000-fold reduction in computation time and near linear scaling with time-series length, despite an average reduction in classification accuracy of just 7\%.
\emph{catch22} captures a diverse and interpretable signature of time series in terms of their properties, including linear and non-linear autocorrelation, successive differences, value distributions and outliers, and fluctuation scaling properties.
We provide an efficient implementation of \textit{catch22}, accessible from many programming environments, that facilitates feature-based time-series analysis for scientific, industrial, financial and medical applications using a common language of interpretable time-series properties.
\end{abstract}

\section{Introduction}
\label{sec:intro}
Time series, ordered sets of measurements over time, enable the study of the dynamics of real-world systems and have become a ubiquitous form of data.
Quantitative analysis of time series is vital for countless domain applications, including in industry (e.g., to detect production irregularities), finance (e.g., to identify fraudulent transactions), and medicine (e.g., to diagnose pathological heartbeat patterns).
As modern time-series datasets have grown dramatically in size, there is a pressing need for efficient methods to solve problems including time-series visualization, clustering, classification, and anomaly detection.

Many applications involving time series, including common approaches to clustering and classification, are based on a defined measure of similarity between pairs of time series.
A straightforward similarity measure---for short, aligned time series of equal length---quantifies whether time-series values are close on average (across time) \cite{Berndt1994UsingSeries,Vlachos2002DiscoveringTrajectories,Moon2001Duality-basedDatabases,Faloutsos1994FastDatabases,Ye2009TimeShapelets}.
This approach does not scale well, often quadratic in both number of time series and series length \cite{Bagnall2017}, due to the necessity to compute distances (often using expensive elastic metrics) between all pairs of objects. 
An alternative approach involves defining time-series similarity in terms of extracted features that are the output of time-series analysis algorithms \cite{Fulcher2014,Fulcher2018Feature-basedAnalysis}.
This approach yields an interpretable summary of the dynamical characteristics of each time series that can then be used as the basis of efficient classification and clustering in a feature space using conventional machine-learning methods.

The number of time-series analysis methods that have been devised to convert a complex time-series data stream into an interpretable set of real numbers is vast, with contributions from a diverse range of disciplinary applications.
Some examples include standard deviation, the position of peaks in the Fourier power spectrum, temporal entropy, and many thousands of others \cite{Fulcher2018Feature-basedAnalysis,Fulcher2013}.
From among this wide range of possible features, selecting a set of features that successfully captures the dynamics relevant to the problem at hand has typically been done manually \cite{Timmer1993CharacteristicsSeries,Nanopoulos2001,Morchen2003,Wang2006,Bagnall2012TransformationClassification.}.
However, subjective feature selection leaves uncertain whether a different feature set may have optimal performance on a task at hand.
Addressing this shortcoming, recent methods have been introduced that take a systematic, data-driven approach involving large-scale comparisons across thousands of time-series features \cite{Fulcher2013,Fulcher2017Hctsa:Extraction}.
This `highly-comparative' approach involves comparison across a comprehensive collection of thousands of diverse time-series features and has recently been operationalized as the \textit{hctsa} (highly comparative time-series analysis) toolbox \cite{Fulcher2013,Fulcher2017Hctsa:Extraction,Fulcher2014}.
\textit{hctsa} has been used for data-driven selection of features that capture the informative properties in a given dataset in applications ranging from classifying Parkinsonian speech signals \cite{Fulcher2013} to identifying correlates of mouse-brain dynamics in the structural connectome \cite{Sethi2017StructuralBrain}.
These applications have demonstrated how automatically constructed feature-based representations of time series can, despite vast dimensionality reduction, yield competitive classifiers that can be applied to new data efficiently \cite{Fulcher2014}.
Perhaps most importantly, the selected features provide interpretable understanding of the differences between classes, and therefore a path towards a deeper understanding of the underlying dynamical mechanisms at play.

Selecting a subset of features from thousands of candidate features is computationally expensive, making the highly-comparative approach unfeasible for some real-world applications, especially those involving large training datasets \cite{Bandara2017ForecastingApproach,10.1007/978-3-030-01771-2_15,Biason2017EC-CENTRIC:Design}.
Furthermore, the \textit{hctsa} feature library requires a Matlab license to run, limiting its widespread adoption.
Many more real-world applications of time-series analysis could be tackled using a feature-based approach if a reduced, efficient subset of features, that capture the diversity of analysis approaches contained in \textit{hctsa}, was developed.

In this study we develop a data-driven pipeline to distill reduced subsets of the most useful and complementary features from thousands of initial candidates, such as those in the \textit{hctsa} toolbox.
Our approach involves scoring the performance of each feature independently according to its performance across a calibration set of 93 time-series classification problems \cite{BagnallTheRepository}.
We show that the performance of an initial (filtered) pool of 4791 features from \textit{hctsa} (mean class-balanced accuracy across all tasks: 77.2\%) can be well summarized by a smaller set of just 22 features (mean accuracy: 71.7\%).
We denote this high-performing subset of time-series features as \textit{catch22} (22 CAnonical Time-series CHaracteristics).
The \textit{catch22} feature set:
(1) computes quickly ($\sim$0.5 second/ 10\,000 samples, roughly a thousand times faster than the full \textit{hctsa} feature set in Matlab) and empirically scales almost linearly, $\mathcal{O}(N^{1.16})$;
(2) provides a low-dimensional summary of time series into a concise set of interpretable characteristics that are generically useful for many real-world time-series datasets; and
(3) is implemented in C with wrappers for python, R, and Matlab, facilitating fast time-series clustering and classification.
We envisage \textit{catch22} expanding the set of problems---including scientific, industrial, financial, and medical applications---that can be tackled using a common feature-based language of canonical time-series properties.

\section{Methods}
\label{sec:methods}

We here describe the datasets we evaluate features on, the time-series features contained in the \textit{hctsa}-toolbox \cite{Fulcher2013,Fulcher2017Hctsa:Extraction} and the selection pipeline to generate a reduced feature subset.

\subsection{Data}
\label{sec:data}

To select a reduced set of useful features, we need to define a measure of usefulness.
Here we use a diverse calibration set of time-series classification tasks from the UEA/UCR (University of East Anglia and University of California, Riverside) Time-Series Classification Repository \cite{Bagnall2017}.
The number of time series per dataset ranges from 28 (`ECGMeditation') to 33\,274 (`ElectricalDevices') adding up to a total of 147\,198 time series.
Individual time series range in length from 24 samples (`ItalyPowerDemand') to 3750 samples (`HeartbeatBIDMC'), and datasets contain between 2 classes (e.g., `BeetleFly') and 60 classes (`ShapesAll').
For 85 of the 93 datasets, unbalanced classification accuracies were provided for different shape-based classifiers such as dynamic time warping (DTW) \cite{Berndt1994UsingSeries} nearest neighbor, as well as for hybrid approaches such as COTE \cite{Bagnall2016Time-seriesEnsembles}.
All unbalanced accuracies, $a^\text{ub}$, were computed using the fixed training-test split provided by the UCR repository, as the proportion of class predictions that matched the actual class labels:
\begin{equation} \label{eq:unbalancedaccuracy}
    a^\text{ub}(y, \hat{y}) = \frac{1}{n_\text{TS}} \sum_{l=1}^{n_\text{TS}} \mathbbm{1}{(\hat{y}_l = y_l)},
\end{equation}
where $y_l$ is the actual class, $\hat{y}_l$ is the predicted class, $n_\text{TS}$ is the total number of time series in the dataset, and $\mathbbm{1}{}$ is the indicator function.

\subsection{Time-series features}
\label{sec:timeseriesfeatures}
Our aim is to obtain a data-driven subset of generically useful time-series features by comparing across as diverse a set of time-series analysis algorithms as possible.
An ideal starting point for such an exercise is the comprehensive library of over 7500 features provided in the \textit{hctsa} toolbox \cite{Fulcher2013,Fulcher2017Hctsa:Extraction}.
Features included in \textit{hctsa} are derived from a wide conceptual range of algorithms, including measurements of the basic statistics of time-series values (e.g., location, spread, Gaussianity, outlier properties), linear correlations (e.g., autocorrelation, power spectral features), stationarity (e.g., StatAv, sliding window measures, prediction errors), entropy (e.g., auto-mutual information, Approximate Entropy, Lempel-Ziv complexity), methods from the physical nonlinear time-series analysis literature (e.g., correlation dimension, Lyapunov exponent estimates, surrogate data analysis), linear and nonlinear model parameters, fits, and predictive power (e.g., from autoregressive moving average (ARMA), Gaussian Process, and generalized autoregressive conditional heteroskedasticity (GARCH) models), and others (e.g., wavelet methods, properties of networks derived from time series, etc.) \cite{Fulcher2013,Fulcher2017Hctsa:Extraction}.
Features were calculated in Matlab 2017a (a product of The MathWorks, Natick, MA) using \textit{hctsa} v0.97.
For each dataset, each feature was linearly rescaled to the unit interval.

We performed an initial filtering of all 7658 \textit{hctsa} features based on their characteristics and general applicability.
Because the vast majority of time series in the UCR/UEA repository are $z$-score normalized\footnote{With the notable exception of four unnormalized datasets: `AALTDChallenge', `ElectricDeviceOn', `ECGMeditation', `HeartbeatBIDMC'.}, we first removed the 766 features that are sensitive to the mean and variance of the distribution of values in a time series based on keywords assigned through previous work \cite{Fulcher2013}, resulting in a set of 6892 features.
We note that on specific tasks with non-normalized data, features of the raw value distribution (such as mean, standard deviation, and higher moments) can lead to significant performance gains and that for some applications, this initial preselection is undesirable \cite{Dau2018UCR2018}. Given a suitable collection of datasets in which the raw value distributions contain information about class differences, our pipeline can easily skip this preselection. 
We next excluded the features that frequently outputted special values, which indicate that an algorithm is not suitable for the input data, or that it did not evaluate successfully.
Algorithms that produced special-valued outputs on at least one time series in more than 80\% of our datasets were excluded: a total of 2101 features (across datasets, minimum: 655, maximum: 3427, mean: 1327), leaving a remaining set of 4791 features.
This relatively high number of features lost during preselection reflects our strict exclusion criterion for requiring real-valued outputs across a diverse range of input data, and the restricted applicability of many algorithms (e.g., that require a minimum length of input data, require positive-valued data, or cannot deal with data repeated identical values).
For example, the datasets with the most special-valued features are `ElectricDevices' (3427 special-valued features), which contains 96-sample time series with many repeated values (e.g., some time series contain just 10 unique values), and `ItalyPowerDemand' (2678 special-valued features), which consists of very short (24-sample) time series.
The 4791 features that survived the preselection gave real-valued outputs on at least 90\% of the time series of all datasets, and 90\% of them succeeded on at least 99\% of time series.

\subsection{Performance-based selection}
\label{sec:featureselection}

\begin{figure*}
  \includegraphics[width=\textwidth]{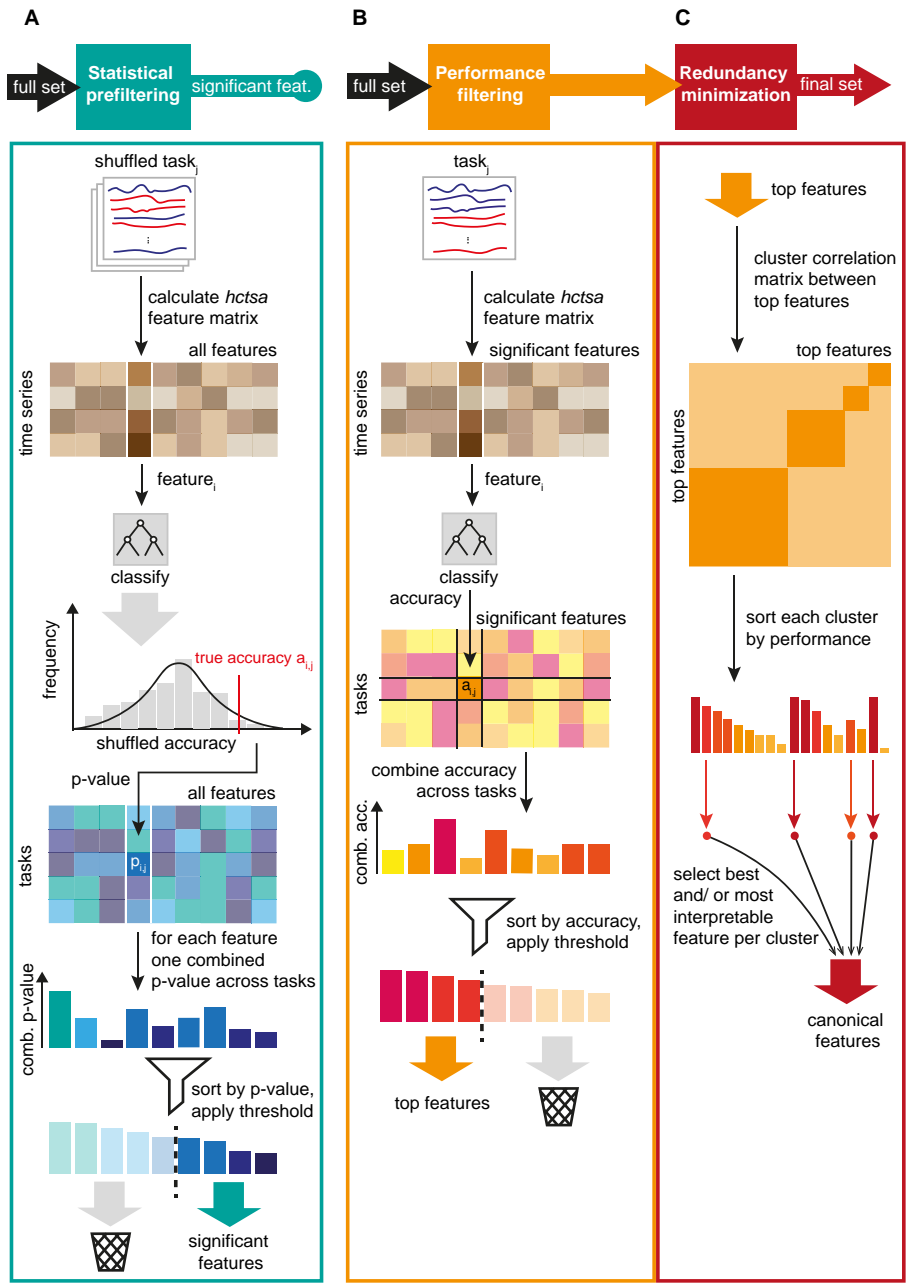}
\caption{\textbf{Given a set of classification tasks, our pipeline selects a reduced set of high-performing features while minimizing inter-feature redundancy.}
\textbf{A} \textit{Statistical prefiltering}: We identified features with performance consistent with that of random number generators.
To this end, we derived null accuracy distributions for each feature on each task by classifying repeatedly on shuffled class labels.
$P$-values from those null distributions were combined across datasets to identify features with performance consistent with random-number generators.
\textbf{B} \textit{Performance filtering}: We selected an intermediate set of top features by ranking and thresholding the combined accuracy over all datasets.
\textbf{C} \textit{Redundancy minimization}: Top performing features were clustered into groups with similar performance across tasks to minimize redundancy between the final set of canonical features.
We selected a single representative feature per cluster to yield a canonical feature set.
}
\label{fig:pipeline}
\end{figure*}

In contrast to typical feature selection algorithms, which look for combinations of features that perform well together, we seek interpretable features that \textit{individually} possess discriminatory power on real world data and are complementary to each other.
To this end we used the pipeline depicted in Fig.~\ref{fig:pipeline}, which evaluates the univariate classification performance of each feature on each task, combines feature-scores across tasks, and then selects a reduced set of generically useful features across a two-step filtering process which involves:
(1) \textit{performance filtering}: select features that perform best across all tasks, and
(2) \textit{redundancy minimization}: reduce redundancy between features.
The method is general and is easily extendable to different sets of classification tasks, or to different initial pools of features.
All analysis was performed in Python 2.7 using \verb|scikit-learn| and code to reproduce all of our analyses is accessible on GitHub (\verb|https://github.com/chlubba/op_importance|).

\subsection{Quantifying feature performance}
\label{sec:featureperformance}

Our pipeline (Fig.~\ref{fig:pipeline}) requires a method to score the performance of individual features across classification tasks.
We scored each feature by its ability to distinguish the labeled classes in each of our $M = 93$ classification tasks and then computed a combined performance score for that feature across all tasks.
Classification was performed using a decision tree with stratified cross validation with $N_\text{CV}$ folds.
The number of folds, $N_\text{CV}$, was chosen separately for each task according to:
\begin{equation} \label{eq:nfolds}
    N_\text{CV} = \min\left\{10, \max\left[2, \min_{k=1}^{N_c}\left(\sum_{l=1}^{N_\text{TS}}\mathbbm{1}{\left(y_l = k\right)}\right)\right]\right\},
\end{equation}
where $N_\text{TS}$ is the number of time series, $N_c$ is the number of classes, and $y_l$ is the class-label of the $l$th time series.

For feature $i$ ($i = 1, ..., 4791$) on classification task $j$ ($j = 1, ..., M$), we computed the mean class-balanced classification accuracy across folds $a_{i,j}$ as a performance score.
\begin{equation}
\label{eq:balancedaccuracy}
    a_{i,j}(y, \hat{y}, w) = \frac{1}{\sum_{l=1}^{N_\text{TS}}{w_l}} \sum_{l=1}^{N_\text{TS}} \mathbbm{1}{(\hat{y}_l = y_l)} w_l,
\end{equation}
where the weights for each time series $w_l$ compensate for imbalances in the number of samples per class, $w_l = 1/\sum_{m=1}^{N_\text{TS}}{\mathbbm{1}{(y_m = y_l)}}$.
To combine scores across tasks, we computed a normalized accuracy of the $j$th task by dividing raw feature accuracies, $a_{i,j}$, by the mean accuracy across all features on that task, $\bar{a}_j$, as follows:
\begin{equation}
\label{eq:accuracynormalisation}
a^\mathrm{n}_{i,j} = \frac{a_{i,j}}{\bar{a}_j}.
\end{equation}
This allowed us to quantify the performance of each feature on a given task relative to the performances of other features in our library.

Finally, the combined feature-accuracy-score across all tasks, $a^\mathrm{n,c}_i$, was calculated as the mean over normalized accuracies, $a^\mathrm{n}_{i,j}$, on our $M = 93$ tasks:
\begin{equation}
\label{eq:accuracycombination}
a^\mathrm{n,c}_{i} = \frac{1}{M}\sum_{j=1}^M a^\mathrm{n}_{i,j}.
\end{equation}


\subsection{Statistical prefiltering}
\label{sec:featuresignificance}

Given the size and diversity of features in \textit{hctsa}, we first wanted to determine whether some features exhibited performance consistent with chance on this set of classification tasks.
To estimate a $p$-value for each feature on each classification task, we generated null accuracy distributions, $a^s_{i,j}$ (superscript $s$ incidating `shuffeled'), using a permutation-based procedure that involved 1000 repeats of the classification procedure using randomly shuffled class labels, shown in Fig.~\ref{fig:pipeline}A.
From visual inspection, the null distributions were mostly unimodal and approximately normally distributed and, as expected, had higher variance on datasets with fewer time series.
Accordingly, we estimated a $p$-value for each feature-task combination by fitting a Gaussian probability distribution to the null accuracy distributions, $a^s_{i,j}$, as the probability of the shuffled accuracy being higher than the observed accuracy. 

To combine $p$-values for a given feature across all classification tasks, we used Fisher's method \cite{Fisher1925StatisticalWorkers} and
corrected for multiple hypothesis testing across features using the Holm-Bonferroni method \cite{Holm1979AProcedure} at a significance level of 0.05.

\subsection{Selecting a canonical set of features}
\label{sec:redundancyremoval}

From the features that performed significantly better than chance, we selected a subset of $\beta$ high-performing features by ranking them by their combined normalized accuracy $a^\mathrm{n,c}$ (Fig.~\ref{fig:pipeline}B), comparing values in the range $100 \leq \beta \leq 1000$.
As shown in Fig.~\ref{fig:pipeline}C, we then aimed to reduce the redundancy in these top-performing features, defining redundancy in terms of patterns of performance across classification tasks.
To achieve this, we used hierarchical clustering on the Pearson correlation distance, $d_{ij} = 1 - r_{ij}$ between the $M$-dimensional performance vectors of normalized accuracies $a^\mathrm{n}$ of features $i$ and $j$.
Clustering was performed using complete linkage at a threshold $\gamma = 0.2$ to form clusters of similarly performing features, that are all inter-correlated by $r_{ij} > 1 - \gamma$ (for all $i,j$ in the cluster).
We then selected a single feature to represent each cluster, comparing two different methods:
(i) the feature with the highest normalized accuracy combined across tasks, and
(ii) manual selection of representative features to favour interpretability (while also taking into account computational efficiency and classification performance).


\subsection{Overall classification performance}
\label{sec:setperformance}

To evaluate the classification performance of different feature sets, and compare our feature-based classification to alternative time-series classification methods, we used two different accuracy measures.
Comparisons between different sets of \textit{hctsa}-features were based on the mean class-balanced accuracy across $M$ tasks and $N_\text{CV}$ cross-validation folds:
\begin{equation} \label{eq:totalbalancedacc}
    a_\text{tot} = \frac{1}{M}\sum_{j=1}^{M}\frac{1}{N_{\text{CV},j}}\sum_{k=1}^{N_{\text{CV},j}}  a_{j,k}.
\end{equation}

When comparing our feature sets to existing methods we used the mean unbalanced classification accuracy across tasks as in Eq.~\eqref{eq:unbalancedaccuracy} on the given train-test split to match the metric used for the accuracies supplied with the UEA/UCR repository:
\begin{equation} \label{eq:totalunbalancedacc}
    a^\text{ub}_\text{tot} = \frac{1}{N_\text{tasks}}\sum_{j=1}^{N_\text{tasks}} a^\text{ub}_{j}.
\end{equation}

\subsection{Execution times and scaling}
\label{sec:executiontime}

One of the merits of a small canonical feature set for time-series characterization is that it is quick to compute.
To compare the execution time of different feature sets, we used a benchmark set of 40 time series from different sources, including simulated dynamical systems, financial data, medical recordings, meteorology, astrophysics, and bird sounds (see Sec.~\ref{sec:empiricalts} for a complete list).
To estimate how features scale with time-series length, we generated multiple versions for each of our 40 reference time series of different lengths from 50 to 10\,000 samples.
Lengths were adapted by either removing points after a certain sample count or by up-sampling of the whole time series to the desired length.
Execution times were obtained on a 2.2\,GHz Intel Core i7, using single-threaded execution (although we note that feature calculation can be parallelized straightforwardly).

\subsection{Selecting the two most informative features from a small subset}
\label{sec:2features}

For the purpose of quickly analyzing a dataset visually in feature space, it can be helpful to identify the two features that, taken together, are the most informative to distinguish between time-series classes of the selected dataset. To this end, we used sequential forward selection \cite{Whitney1971ASelection,Fulcher2014} that first selects a single feature which achieves the best mean class-balanced accuracy across cross-validation folds and then iterates over the remaining features to select the one that, combined with the first feature, reaches the best accuracy.



\section{Results}
\label{sec:results}

We present results of using our pipeline to obtain a canonical set of 22 time-series features from an initial pool of 4791 candidates.
We name our set \textit{catch22} (22 CAnonical Time-series CHaracteristics), which approximates the classification performance of the initial feature pool to 90\% and computes in less than 0.5\,s on 10\,000 samples.



\subsection{Performance diversity across classification tasks}

\begin{figure*}
  \includegraphics[width=\textwidth]{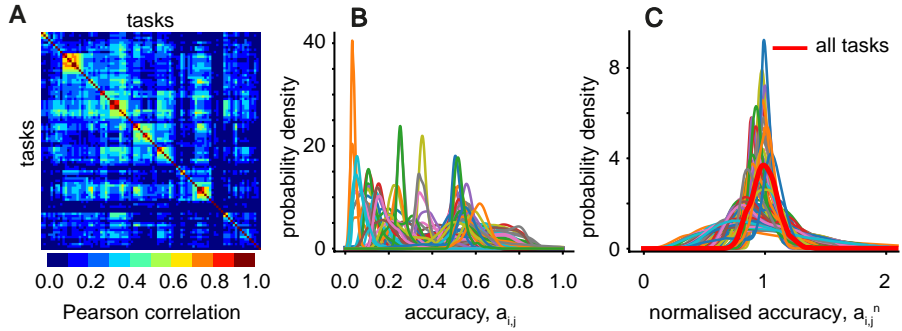}
\caption{
\textbf{Normalization of feature accuracies allows comparison of performance scores across a diverse set of 93 classification tasks.
}
\textbf{A}~A $93 \times 93$ matrix of Pearson correlation coefficients between the 4791-dimensional accuracy vectors of pairs of tasks, reordered according to hierarchical linkage clustering.
\textbf{B}~Each line shows the smoothed distribution over feature-accuracies on one classification task.
Differences in task difficulty are reflected by a wide range of accuracies.
\textbf{C}~The accuracies plotted in \textbf{B} were normalized by the mean accuracy of each task, as in Eq.~\eqref{eq:accuracynormalisation}.
The red line shows the distribution of normalized and combined accuracies across all tasks, Eq.~\eqref{eq:accuracycombination}.
}
\label{fig:tasks}
\end{figure*}

We first analyze the 93 classification tasks, which are highly diverse in their properties (see Sec.~\ref{sec:data}) and difficulty, as shown in Fig.~\ref{fig:tasks}.
We characterized the similarity of two tasks in terms of the types of features that perform well on them, as the Pearson correlation coefficient between accuracies of all features, shown in Fig~\ref{fig:tasks}A.
The figure reveals a diversity of performance signatures across tasks: for some groups of tasks, similar types of features contribute to successful classification, whereas very different types of features are required for other tasks.
The 93 tasks also vary markedly in their difficulty, as judged by the distribution of accuracies, $a_{i,j}$, across tasks, shown in Fig.~\ref{fig:tasks}B.
We normalized feature accuracies across tasks by dividing them by the mean accuracy of the task at hand, Eq.~\eqref{eq:accuracynormalisation}, yielding normalized accuracies, $a_{i,j}^\mathrm{n}$, that were comparable across tasks, shown in Fig.~\ref{fig:tasks}C.
Note that this normalized accuracy scores features relative to all other features on a given task. 
The red line in Fig.~\ref{fig:tasks}C shows the distribution of the mean normalized accuracies across tasks $a_i^{n,c}$, Eq.~\eqref{eq:accuracycombination}.

\subsection{Features with performance consistent with chance}

To detect whether some features in \textit{hctsa} exhibit a combined performance across classification tasks that is consistent with the performance of a random-number generator, we used a permutation-testing based procedure (described in Sec.~\ref{sec:featuresignificance}).
At a significance level $p < 0.05$, 145 of the 4791 features (or 3\%) exhibited chance-level performance.
These 145 features (listed in Supplementary Table~\ref{tab:insignificantops}) were mostly related to quantifying complex dynamics in longer time series, such as nonlinear time-series analysis methods, long-range automutual information; properties that are not meaningful for the short, shape-based time-series patterns that dominate the UEA/UCR database.

\subsection{Top-performing features}

\begin{figure*}
  \includegraphics[width=\textwidth]{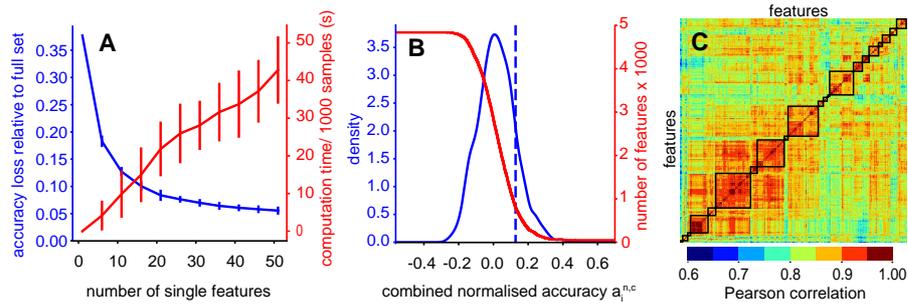}
\caption{
\textbf{The mean classification performance of the full feature set can be well approximated (to within 10\%) by as few as 20 features.}
\textbf{A}~While computation time is observed to rise linearly with increasing number of single selected features, the relative difference in accuracy to the full set of 4791 features starts to saturate at around 20 features.
The performance of our final set of features is not highly sensitive to the size of our intermediate set of top-performing features, $\beta$.
Error bars signify standard deviation over both relative loss in accuracy and computation time for different numbers of top features ($\beta = 100, 200, ..., 1000$), which were clustered to obtain single features (see Methods Sec.~\ref{sec:redundancyremoval}).
\textbf{B}~We select the number of top features from a relative threshold on the combined normalized accuracy across tasks shown as a dashed blue vertical line, yielding a set of 710 high-performing features.
\textbf{C}~High-performing features were clustered on performance-correlation distances using hierarchical complete linkage clustering, using a distance threshold $\gamma$ of of 0.2, yielding 22 clusters.
}
\label{fig:redundancy}
\end{figure*}

As a second step in our pipeline, we ranked features by their combined accuracy across tasks and then selected a subset of $\beta$ best performers.
How important is the choice of $\beta$?
Fig.~\ref{fig:redundancy}A shows how the relative difference in classification accuracy between full and reduced set $(a_{tot,full}-a_{tot,subset})/a_{tot,full}$ (blue line) and computation time (red line) evolve when increasing the number of clusters ($1$ to $50$) into which the top performing features are grouped. Error bars signify the standard deviation over accuracies and computation times when starting from different numbers of top performers $\beta = 100, 200, 300, ..., 1000$. Their tightness demonstrates that the accuracy of the final feature subset was not highly sensitive to the value of $\beta$. Computation time is more variable.
Fig.~\ref{fig:redundancy}A further shows that the relative difference in class-balanced accuracy between full and reduced sets saturated under 10\% for between 20 and 30 selected features.
To obtain a reduced set of high-performing features, we used a threshold on the combined normalized accuracy $a^{n,c}$ of one standard deviation above the mean, $a_\mathrm{th} = \overline{a^\mathrm{n,c}} + \sigma_{a^\mathrm{n,c}}$, shown in Fig.~\ref{fig:redundancy}B, yielding a set of 710 features.

\subsection{A canonical feature set, \textit{catch22}}

We reduced inter-feature redundancy in \textit{hctsa} \cite{Fulcher2013}, by applying hierarchical complete linkage clustering based on the correlation distances between performance vectors of the set of 710 high-performing features, as shown in Fig.~\ref{fig:redundancy}C.
Clustering at a distance threshold $\gamma = 0.2$ (see Sec.~\ref{sec:redundancyremoval}) yielded 22 clusters of similarly-performing features, where the correlation of performance vectors between all pairs of features within each cluster was greater than 0.8.
Different values of $\gamma$ correspond to different penalties for redundancy; e.g., higher values ($\gamma > 0.4$) group all features into a single cluster, whereas low values would form many more clusters and increase the size and complexity of computing the resulting canonical feature set.
We found $\gamma = 0.2$ to represent a good compromise that yields a resulting set of 22 clusters that matches the saturation of performance observed between 20--30 features (Fig.~\ref{fig:redundancy}A).

\begin{figure*}
  \includegraphics[width=\textwidth]{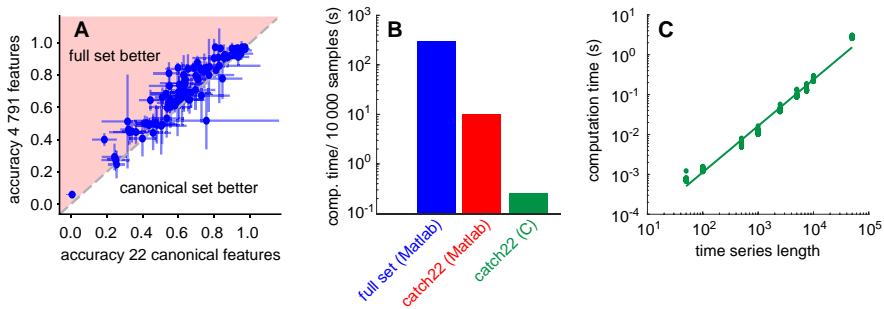}
\caption{
\textbf{The \textit{catch22} set of 22 features approximates the classification performance of all 4791 features despite a dramatic reduction in computation time.}
\textbf{A}~Each point represents one dataset in its balanced accuracy based on the \textit{catch22} feature set ($x$-axis) and the full set of 4791 features ($y$-axis).
Error bars signify standard deviation across cross-validation folds.
\textit{catch22} performs only a relative 7.5\% worse than the full set of 4791 features: 71.7\% vs 77.2\% mean class-balanced accuracy across tasks $a^\text{tot}$ as defined in Eq.~\eqref{eq:totalbalancedacc}.
\textbf{B}~Bars represents average over serial computation times for each of our 40 reference time series at a length of 10\,000 samples using the full set of 4791 features, \textit{catch22} in Matlab and \textit{catch22} in C.
From full set in Matlab to \textit{catch22} in C, computation time decreases from $\sim$300 seconds to less than 0.5\,s.
\textbf{C}~Each dot shows computation time for one of the 40 reference time series adjusted to different lengths for the C-implemented \textit{catch22} set.
The linear fit in the logarithmic plot reveals an almost linear scaling, with a scaling exponent of 1.16.
See Sec.~\ref{sec:executiontime} for a description of the data.
}
\label{fig:performance}
\end{figure*}

We next aimed to capture the behavior of each of the 22 clusters as a single feature with the most representative behavior of its cluster.
We first achieved this automatically: selecting the feature with the highest combined normalized accuracy $a_í^{n,c}$ from each cluster.
When classifying our tasks with this set of 22 best estimators, it reached an overall class-balanced accuracy over folds and tasks $a^\text{tot}$, Eq.~\eqref{eq:totalunbalancedacc}, of $\sim$70\%, compared to $\sim$77\% using the full set.
However, it is desirable for our 22 features to be as fast and easily interpretable as possible.
For 6 of the 22 clusters, the top-performing feature was relatively complicated to compute and only offered a relatively small improvement in performance relative to simpler features with similar performance in the same cluster.
In these cases, we manually selected a simpler and more interpretable feature, yielding a final canonical set of 22 features which we call \textit{catch22} (CAnonical Time-series CHaracteristics).
The 22 features that make up \textit{catch22} are described in Tab.~\ref{tab:canonicalfeatures}.
The \textit{catch22} features reflect the diverse and interdisciplinary literature of time-series analysis methods that have been developed to date \cite{Fulcher2013}, simultaneously probing different types of structure in the data, including properties of the distribution of values in the time series, its linear and non-linear autocorrelation, predictability, scaling of fluctuations, and others.

\clearpage
\thispagestyle{empty}
\begin{landscape}
\begin{table}[ht]
        \centering 
        \begin{tabular}{lr} 
        \toprule
        \textit{hctsa} feature name & Description\\ 
        \midrule 
\textit{Distribution} & \\
\verb|DN_HistogramMode_5| & Mode of $z$-scored distribution (5-bin histogram)\\
\verb|DN_HistogramMode_10| & Mode of $z$-scored distribution (10-bin histogram)\\
\textit{Simple temporal statistics} & \\
\verb|SB_BinaryStats_mean_longstretch1| & Longest period of consecutive values above the mean\\
\verb|DN_OutlierInclude_p_001_mdrmd| & Time intervals between successive extreme events above the mean\\
\verb|DN_OutlierInclude_n_001_mdrmd| & Time intervals between successive extreme events below the mean\\

\textit{Linear autocorrelation} & \\
\verb|CO_f1ecac| & First $1/e$ crossing of autocorrelation function\\
\verb|CO_FirstMin_ac| & First minimum of autocorrelation function\\
\verb|SP_Summaries_welch_rect_area_5_1| & Total power in lowest fifth of frequencies in the Fourier power spectrum\\
\verb|SP_Summaries_welch_rect_centroid| & Centroid of the Fourier power spectrum\\
\verb|FC_LocalSimple_mean3_stderr| & Mean error from a rolling 3-sample mean forecasting\\

\textit{Nonlinear autocorrelation} & \\
\verb|CO_trev_1_num| & Time-reversibility statistic, $\langle(x_{t+1}-x_t)^3\rangle_t$\\
\verb|CO_HistogramAMI_even_2_5| & Automutual information, $m=2, \tau=5$\\
\verb|IN_AutoMutualInfoStats_40_gaussian_fmmi| & First minimum of the automutual information function\\

\textit{Successive differences} & \\
\verb|MD_hrv_classic_pnn40| & Proportion of successive differences exceeding $0.04\sigma$ \cite{Mietus2002TheMeasure}\\
\verb|SB_BinaryStats_diff_longstretch0| & Longest period of successive incremental decreases\\
\verb|SB_MotifThree_quantile_hh| & Shannon entropy of two successive letters in equiprobable 3-letter symbolization\\
\verb|FC_LocalSimple_mean1_tauresrat| & Change in correlation length after iterative differencing\\
\verb|CO_Embed2_Dist_tau_d_expfit_meandiff| & Exponential fit to successive distances in 2-d embedding space\\

\textit{Fluctuation Analysis} & \\
\verb|SC_FluctAnal_2_dfa_50_1_2_logi_prop_r1| & Proportion of slower timescale fluctuations that scale with DFA (50\% sampling) \\
\verb|SC_FluctAnal_2_rsrangefit_50_1_logi_prop_r1| &  Proportion of slower timescale fluctuations that scale with linearly rescaled range fits\\

\textit{Others} & \\
\verb|SB_TransitionMatrix_3ac_sumdiagcov| & Trace of covariance of transition matrix between symbols in 3-letter alphabet\\
\verb|PD_PeriodicityWang_th0_01| & Periodicity measure of \cite{Wang2007Structure-basedClustering}\\
\bottomrule
\end{tabular}
\captionof{table}{
\textbf{The \textit{catch22} feature set spans a diverse range of time-series characteristics representative of the diversity of interdisciplinary methods for time-series analysis.}
Features in \textit{catch22} capture time-series properties of the distribution of values in the time series, linear and nonlinear temporal autocorrelation properties, scaling of fluctuations, and others.
}
        \label{tab:canonicalfeatures}
\end{table}
\end{landscape}

Using the diverse canonical \textit{catch22} feature-subset, the mean class-balanced accuracy across all datasets, $a^\text{tot}$, of \textit{catch22} was $\sim$72\%, a small reduction relative to the $\sim$77\% achieved when computing all 4791 features. See Fig.~\ref{fig:performance}A for a dataset-by-dataset scatter.
The change in mean accuracy across folds and tasks, $a^\text{tot}$, from using the 22 features of \textit{catch22} instead of all 4791 features cote the properties of a given dataset, but there was an average reduction in class-balanced accuracy (mean across folds) of 7.5\% relative to the full set accuracy (77.2\% full vs 71.7\% canonical, 7.5 percentage points).
For some difficult problems, the increased computational expense of the full set of 4791 features yields a large boost in classification accuracy (accuracy of \textit{catch22} lower by a relative difference of 37\% for the dataset `EthanolLevel'; 50.2\% full vs 31.8\% \textit{catch22}).
The reduced set gave better mean performance in only a small number of cases: e.g., for `ECGMeditation' with 60\% full vs 81.2\% \textit{catch22}; given that this dataset contained just 28 time series and had a high standard deviation in accuracies of the full set between folds (35.3\%), the performance might not be significantly increased.

How does the performance of the data-driven features, \textit{catch22}, compare to the manually-curated 16-feature set in the \verb|tsfeatures| package \cite{Hyndman2016}?
Reassuringly, the class-balanced accuracies of both feature sets were very similar across the generic UCR/UAE datasets, with a Pearson correlation coefficient $r = 0.93$ (Fig.~\ref{fig:hyndman}).
The mean accuracy across tasks and folds, $a^\text{tot}$, was slightly higher for \textit{catch22} (71.7\%) than \verb|tsfeatures| (69.4\%).
Our pipeline is general, and can select informative subsets of features for any collection of problems; e.g., for a more complex set of time-series classification tasks, our pipeline may yield estimators of more distinctive and complex dynamics.

\begin{figure*}
  \includegraphics[width=\textwidth]{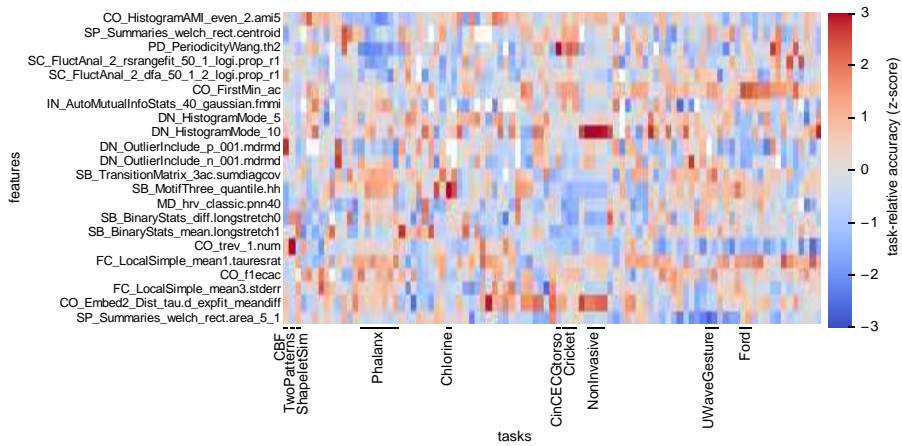}
\caption{
\textbf{The canonical features in \textit{catch22} are sufficiently diverse to enable high performance across diverse classification tasks.}
The matrix shows class-balanced accuracies, $z$-scored per task (column), truncated at $\pm$3, and was reordered by hierarchical linkage clustering based on correlation distance in both columns (93 classification tasks) and rows (22 features).
Similar columns are visible for datasets of the same type.
The \textit{catch22} features each show strengths and weaknesses, and their diversity allows them to complement each other across a range of tasks.
}
\label{fig:performancePerDataset}
\end{figure*}

How diverse are the features in \textit{catch22}? Fig.~\ref{fig:performancePerDataset} displays the class-balanced accuracies of each of the \textit{catch22} features (rows) on each task (columns), z-normalized by task.
Some groups of tasks recruit the same types of features for classification (reflected by groups of columns with similar patterns).
Patterns across rows capture the characteristic performance signature of each feature, and are visually very different, reflecting the diversity of features that make up \textit{catch22}.
This diversity is key to being able to probe the different types of temporal structure required to capture specific differences between labeled classes in different time-series classification tasks in the UCR/UAE repository.
Feature-performances often fit the known dynamics in the data, e.g., for the two datasets `FordA' and `FordB' in which manual inspection reveals class differences in the low frequency content, the most successful feature is `\verb|CO_FirstMin_ac|' which finds the first minimum in the autocorrelation function.
In some datasets, high performance can be attained using just a single feature, e.g., in `ChlorineConcentration' (`\verb|SB_motifThree_quantile.hh|', 52.3\% vs 67.5\% class-balanced mean accuracy over folds $a$ for \textit{catch22} vs all features) and `TwoPatterns' (`\verb|CO_trev_1.num|', 73.4\% vs 88.1\%).

\subsection{Computation time and complexity}

The classification performance using all 4791 features is well approximated by the 22 features in \textit{catch22}, but how much computational effort does it save? 
To maximize the acceleration in execution time and make our condensed subset accessible from all major ecosystems used by the data-mining community, we implemented all \textit{catch22} features in C and wrapped them for R, Python and Matlab. All code is accesible on GitHub \verb|https://github.com/chlubba/catch22|.
Using this C-implementation, the \textit{catch22} feature set can be computed sequentially on all 93 datasets of the UEA/UCR repository in less than 15 minutes on an Intel Core i7. 
On average, the features for each dataset were calculated within 9.4\,s, the slowest being `StarLightCurves' with 97\,s due to its many (9236) relatively long (1024 samples) time series.
The 27 quickest datasets stayed below 1\,s in computation time; the three quickest, `BirdChicken', `Coffee', and `BeetleFly' took less than 0.25\,s.

While time series contained in the UEA/UCR repository are usually short, with an average length of 500 samples, real-world recordings can be substantially longer. 
Therefore, to understand how the computation times of our feature set scale with time-series lengths above those available in the UEA/UCR repository, we used a set of 40 reference time series from diverse sources (described in Sec.~\ref{sec:executiontime}) to evaluate execution times of all \textit{hctsa}- and the \textit{catch22}-features for longer time series.
Fig.~\ref{fig:performance}B shows execution times of different feature sets as a mean over our 40 reference time series at length 10\,000.
The Matlab implementation of \textit{catch22} accelerates computation time by a factor of $\sim$30 compared to the full set of 4791 from $\sim$300s to $\sim$10s.
The C-implementation of \textit{catch22} again reduces execution time by a factor of approximately 30 compared to the Matlab implementation to $\sim$0.3\,s at 10\,000 samples, signifying an approximately 1000-fold acceleration compared to the full \textit{hctsa} feature set in Matlab.
The C-version of \textit{catch22} exhibits near-linear computational complexity, $\mathcal{O}(N^{1.16})$, as shown in Fig.~\ref{fig:performance}C.
Features varied markedly in their execution time, ranging from (C-implemented) \verb|DN_HistogramMode_10| ($<0.1$\,ms for our 10\,000-sample reference series) to \verb|PD_PeriodicityWang_th0_01| (79\,ms), with the latter representing approximately one third of the total computation time for \textit{catch22}.
A further acceleration by a factor of 3 could be achieved through parallelization, limited by the slowest feature \verb|PD_PeriodicityWang_th0_01| which takes up one third of the overall computation time.

\subsection{Performance comparison}

\begin{figure*}
\centering
  \includegraphics[width=12cm]{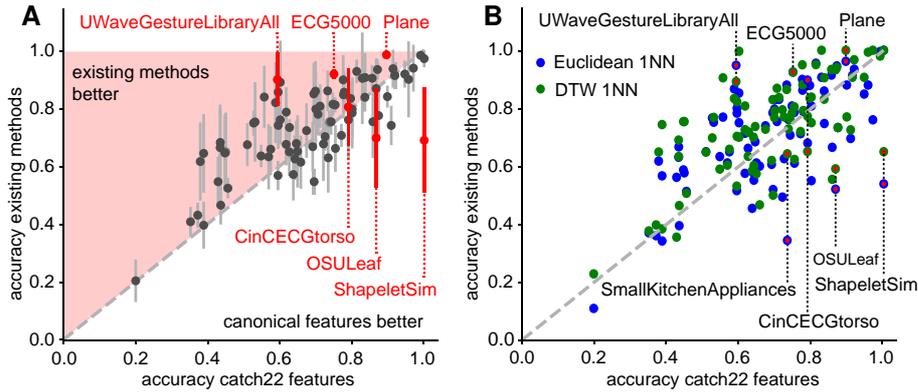}
\caption{
\textbf{Despite massive dimensionality reduction to 22 features, the \textit{catch22} representation often achieves similar or better performance on time-series classification tasks.}
\textbf{A}~Classification accuracy is plotted from using the feature-based \textit{catch22} representation versus the performance of a range of existing methods across the 93 tasks in the UEA/UCR repository.
Each dot represents the mean accuracy of alternative classifiers on a given dataset; error bars show the standard deviation over the 36 considered other methods containing simple full sequence shape-based approaches, over ensembles, shapelets, intervals, to complex transformation ensembles.
An equality gray-dashed line is plotted, and regions in which \textit{catch22} or other methods perform better are labeled.
\textbf{B}~The two purely shape-based classifiers, Euclidean (blue circles) and DTW (green circles) 1 nearest-neighbor, are compared against \textit{catch22} features and a classification tree.
All accuracies are unbalanced, as Eq.~\eqref{eq:unbalancedaccuracy}, and evaluated on the fixed train-test split provided in the UEA/UCR repository.
}
\label{fig:benchmark}
\end{figure*}

Compared to conventional shape-based time-series classifiers, that use distinguishing patterns in the time domain as the basis for classification \cite{Fulcher2014,Fulcher2018Feature-basedAnalysis}, feature-based representations can reduce high-dimensional time series down to a compact and interpretable set of important numbers, constituting a dramatic reduction in dimensionality.
While this computationally efficient representation of potentially long and complex streams of data is appealing, important information may be lost in the process, resulting in poorer performance than alternative methods that learn classification rules on the full time-series object.
To investigate this, we compared the classification performance of \textit{catch22} (using a decision tree classifier as for every classification, see Sec.~\ref{sec:featureperformance}) to that of 36 other classifiers (accuracies obtained from the UEA/UCR repository \cite{Bagnall2017}) including shape-based approaches like Euclidean or DTW nearest neighbor, ensembles of different elastic distance metrics \cite{Lines2015TimeMeasures}, interval methods, shapelets \cite{Ye2009TimeShapelets}, dictionary based classifiers, or complex transformation ensemble classifiers that combine multiple time-series representations (COTE) \cite{Bagnall2016Time-seriesEnsembles}. 
All comparisons are based on (class unbalanced) classification accuracies $a^\text{ub}_\text{tot}$ on a fixed train-test split obtained from UCR/UEA classification repository.
As shown in Fig.~\ref{fig:benchmark}A, most datasets exhibit similar performance between the alternative methods and \textit{catch22}, with a majority of datasets exhibiting better performance using existing algorithms than \textit{catch22}.
However, despite drastic dimensionality reduction, our feature-based approach outperforms the existing methods on a range of datasets, some of which are labeled in Fig.~\ref{fig:benchmark}A.
To better understand the strengths and weaknesses of our low-dimensional feature-based representation of time series, we compared it directly to two of the most well-studied and purely shape-based time-series classification methods: Euclidean-1NN and DTW-1NN (`DTW-R1-1NN' in the UEA/UCR repository), as shown in Fig.~\ref{fig:benchmark}B.
There is an overall high correlation in performance across datasets, with a range of average performance (unbalanced classification rate on the given train-test partition $a^\text{ub}_\text{tot}$): \textit{catch22} (69\%), Euclidean 1-NN (71\%), and DTW 1-NN (74\%).
The most interesting datasets are those for which one of the two approaches (shape-based or feature-based) markedly outperforms the other, as in these cases there is a clear advantage to tailoring your classification method to the structure of the data \cite{Fulcher2018Feature-basedAnalysis}; selected examples are annotated in Fig.~\ref{fig:benchmark}B).
We next investigate the characteristics of time-series datasets that make them better suited to different classification approaches.

\subsection{Characteristics of datasets that favor feature- or shape-based representations}

There is no single representation that is best for all time-series datasets, but rather, the optimal representation depends on the structure of the dataset and the questions being asked of it \cite{Fulcher2018Feature-basedAnalysis}.
In this section we characterize the properties of selected datasets that show a strong preference for either feature-based or shape-based classification, as highlighted in Fig.~\ref{fig:benchmark}.

\begin{figure*}
\centering
  \includegraphics[width=\textwidth]{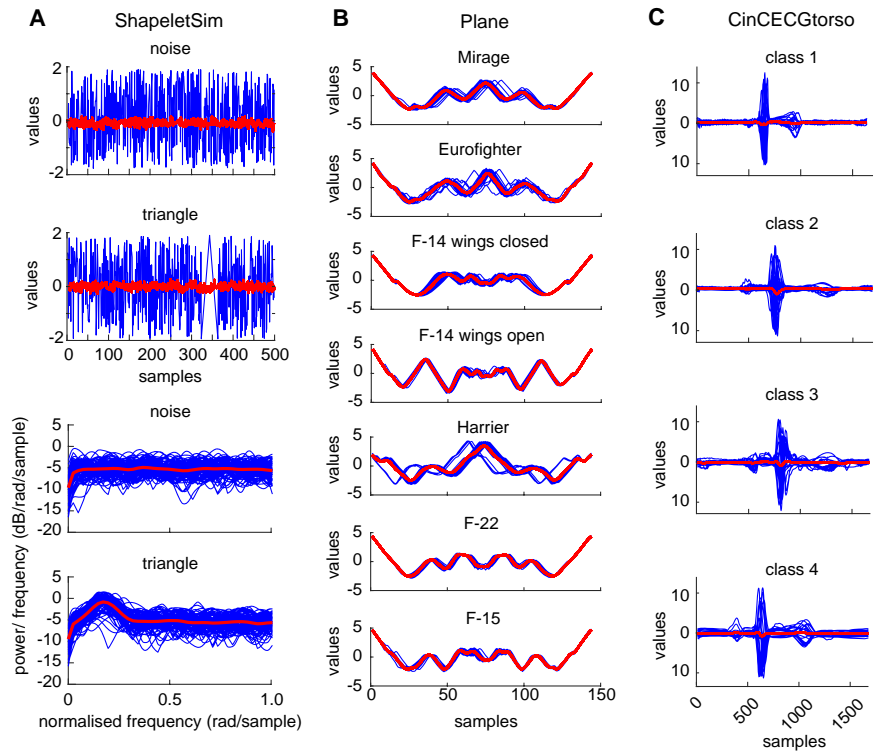}
\caption{
\textbf{Differences in the frequency domain are better picked up by features; subtle differences in shape are better detected by shape-based methods.}
Each subplot represents a class, blue lines show individual time series, red lines show an average over all time series in one class.
\textbf{A}~In the time domain (upper two plots), we show one example time series of the `ShapeletSim' dataset (blue) and the average across all time series (red) for each class.
The lower two plots display the Welch spectra of all time series individually in blue and the average over single time-series spectra in red.
The mean spectra of the two classes differ visibly while there is no reliable difference in the time domain.
\textbf{B}~The individual (blue) and averaged (red) time series of the dataset `Plane' should favor shape-based comparisons because of the highly reliable and aligned shapes in each class.
\textbf{C}~For the dataset `CincECGtorso', all four classes can be well distinguished by their temporal offsets.
}
\label{fig:casestudies}
\end{figure*}

One striking example, is that of `ShapeletSim', where the two labeled classes are much more accurately distinguished using the \textit{catch22} feature-based representation (unbalanced accuracy $a^\text{ub}$ of 100\%) than by all but two existing methods (BOSS \cite{Schafer2015TheNoise} and Fast Shapelets \cite{Rakthanmanon2013FastShapelets}) with a mean and standard deviation over all other classifiers of 69.0 $\pm$ 18.7\% (DTW-1NN 65\%, Euc-1NN 53.9\%).
To understand the discrepancy, we visualized the data in the time-domain, as shown in Fig.~\ref{fig:casestudies}A (upper), where one example time series and the mean in each class are plotted, revealing no consistent time-domain shape across the 100 instances of each class.
However, the two classes of time series are clearly distinguished by their frequency content, as shown in the corresponding Welch power spectra in Fig.~\ref{fig:casestudies}A (lower).
The features in \textit{catch22} capture the temporal autocorrelation properties of each time series in various ways, facilitating an efficient representation to successfully capture class differences in `ShapeletSim'; these differences cannot be captured straightforwardly from the time-domain representation.
Similar datasets without reliable shape differences between classes that exhibit superior performance from \textit{catch22} are e.g., `USOLeaf'  (86.7 \textit{catch22} vs 69.5$\pm$13.3\% others; DTW-1NN 59.1\%, Euc-1NN 52.1\%), `SmallKitchenAppliances' (73.3\% vs 63.3 $\pm$ 12.1\%; DTW-1NN 64.3\%, Euc-1NN 34.4\%).


An example of a dataset that is well-suited to shape-based classification is the seven-class `Plane' dataset, shown in Fig.~\ref{fig:casestudies}B.
Apart from a minority of anomalous instances in e.g., the `Harrier' class, each class has a subtle but robust shape, and these shapes are phase-aligned, allowing shape-based classifiers to accurately capture class differences.
Despite being visually well-suited to shape-based classification, \textit{catch22} captures the class differences with only a small reduction in accuracy $a^\text{ub}$ (89.5\%) compared to the shape-based classifiers (99.2 $\pm$ 1.4\% over all given classifiers; DTW-1NN 100\%, Euc-1NN 96.1\%), demonstrating that feature-based representations can be versatile in capturing differences in time-series shape, despite a substantial reduction in dimensionality.


As a final example we consider the four classes of the `CinCECGtorso' dataset, which are similarly accurately classified by our \textit{catch22} feature-based method (78.9\%) and the average existing classifier (81.3 $\pm$ 13.3\%).
Interestingly, when comparing selected shape-based classifiers in Fig.~\ref{fig:benchmark}, Euclidean-1NN (89.7\%) outperforms the more complex DTW-1NN (65.1\%).
This difference in performance is due to the subtle differences in shape (particularly temporal offset of the deviation from zero) between the four classes, as shown in Fig.~\ref{fig:casestudies}B.
Simple time-domain distance metrics like Euclidean-1NN will capture these important differences well, whereas elastic distance measures like DTW shadow the informative temporal offsets.
Converting to our feature-based representation discards most of the phase-information but still leads to a high classification accuracy.




\subsection{Informative features provide understanding}
\label{sec:informativeFeatureSpaces}

\begin{figure}[ht]
\centering
  \includegraphics[width=\textwidth]{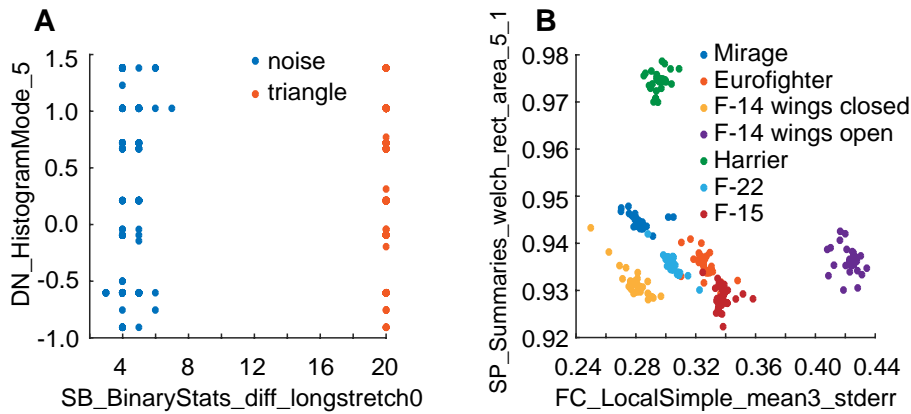}
\cprotect\caption{\textbf{Class differences can be interpreted using feature-based representations of time series.}
We plot a projection of time series into an informative two-dimensional feature space (estimated from \textit{catch22} using sequential forward selection, see Sec.~\ref{sec:2features}), where each time series is a point in the space and colored by its class label.
Plots are shown for two datasets: \textbf{A} `ShapeletSim', and \textbf{B} `Plane'; in both cases, all labeled classes are clearly distinguished in the space.
In `ShapeletSim', \verb|SB_BinaryStats_diff_longstretch0|, which calculates the length of the longest run of consecutive decreases in the time series.
The two features selected for the `Plane' dataset are the local predictability measure, \verb|FC_LocalSimple_mean3_stderr|, and the low-frequency power estimate, \verb|SP_Summaries_welch_rect_area_5_1|.
}
\label{fig:2Dprojection}
\end{figure}

Concise, low-dimensional summaries of time series, that exploit decades of interdisciplinary methods development for time-series analysis, are perhaps most important for scientists because they provide a means to understand class differences.
Often a researcher will favor a method that provides interpretable understanding that can be used to motivate new solutions to a problem, even if it involves a small drop in classification accuracy relative to an opaque, black-box method.
To demonstrate the ability of \textit{catch22} to provide understanding into class difference, we projected all datasets into a two-dimensional feature space determined using sequential forward selection  \cite{Whitney1971ASelection} as described in Sec.~\ref{sec:2features}. 
Two examples are shown in Fig.~\ref{fig:2Dprojection}.
In the dataset `ShapeletSim' (Fig.~\ref{fig:2Dprojection}A), the simple feature, \texttt{SB\_\allowbreak BinaryStats\_\allowbreak diff\_\allowbreak longstretch0}, clearly distinguishes the two classes.
This simple measure quantifies the length of the longest continued descending increments in the data which enables a perfect separation of the two classes because time series of the `triangle' class vary on a slower timescale than `noise' time series.

In the most accurate two-dimensional feature space for the 7-class `Plane' dataset, shown in Fig.~\ref{fig:2Dprojection}B, each class occupies a distinctive part of the space.
The first feature, \texttt{FC\_\allowbreak LocalSimple\_\allowbreak mean3\_\allowbreak stderr} captures variability in residuals for local 3-sample mean predictions of the next datapoint applied to through time, while the second feature, \texttt{SP\_\allowbreak Summaries\_\allowbreak welch\_\allowbreak rect\_\allowbreak area\_5\_1}, captures the proportion of low-frequency power in the time series.
We discover, e.g., that time series of `F-14 wings open' are less predictable from a 3-sample running mean than other planes, and that time series of `Harrier' planes exhibit a greater proportion of low-frequency power than other types of planes.
Thus, in cases when both shape-based and feature-based methods exhibit comparable performance (unbalanced accuracies $a^\text{ub}$ on given split: 89.5\% by \textit{catch22} vs 99.1\% mean over other classifiers), the ability to understand class differences can be a major advantage of the feature-based approach.

\section{Discussion}

Feature-based representations of time-series can distill complex time-varying dynamical patterns into a small set of interpretable characteristics that can be used to represent the data for applications like classification and clustering.
Most importantly, features connect the data analyst to deeper theory, allowing interpretation of the properties of the data that facilitate successful performance.
While large feature libraries have helped to overcome the limitations of manual, subjective curation of time-series features, they are inefficient and computationally expensive.
Overcoming this limitation, here we introduce a methodology to generate small, canonical subsets of features that each display high performance across a given ensemble of tasks, and exhibit complementary performance characteristics with each other.
We apply the method to a set of 93 classification tasks from the UCR/UAE repository, showing how a large library of 4791 features can be reduced to a canonical subset of just 22 features, \textit{catch22}, which displays similar classification accuracy as the full set (relative reduction of 7.5\% on average, 77.2\% vs 71.7\%), computes quickly ($<$0.5s/10\,000 samples), scales approximately linearly with time-series length ($\mathcal{O}(N^{1.16})$), and allows the investigator to learn and understand what types of dynamical properties distinguish the labeled classes of their dataset.
Compared to shape-based methods like dynamic time warping (DTW), \textit{catch22} gives comparable, and often superior classification performance, despite substantial dimensionality reduction.
Using case studies, we explain why some datasets are better suited to shape-based classification (e.g., there are characteristic aligned shapes within each class), while others are better suited to feature-based classification (e.g., where classes do not have a characteristic, temporally aligned shape, but have characteristic dynamical properties that are encapsulated in one or more time-series features).


While some applications may be able to justify the computational expense of searching across a large feature library such as \textit{hctsa} \cite{Fulcher2014,Fulcher2017Hctsa:Extraction}, the availability of an efficient, reduced set of features, as \textit{catch22} 
will make the advantages of feature-based time-series analysis more widely accessible. The methodological pipeline to generate reduced sets has value in itself and was e.g. used to generate a slightly different feature subset for the self-organizing time-series database for data-driven interdisciplinary collaboration \textit{CompEngine} \cite{Fulcher2018CompEngine:Data}.
Unlike the Matlab-based \textit{hctsa}, \textit{catch22} does not require a commercial license to run, computes efficiently, and scales approximately linearly with time-series length in the cases we tested.
This makes it straightforwardly applicable to much longer time series than are typically considered in the time-series classification literature, e.g., for a 10\,000-sample time series, \textit{catch22} computes in 0.5\,s.
As well as being suitable for long recordings, feature-based representations do not require all time series to be the same length (unlike conventional shape-based classifiers), opening up the feature-based approach to new types of datasets.
To facilitate its adoption, we provide an efficient C-implementation of \textit{catch22}, with wrappers for Matlab, Python, and R.

We have shown that the most useful representation of a time series varies widely across datasets, with some problems better suited to feature-based classification, and others better suited to shape-based classification.
The 22 features selected here are tailored to the properties of the UCR/UEA datasets (which are typically short and phase-aligned), but the method we present here is general and could be used to generate reduced feature sets tailored to any application domain of interest that allows individual features to be assigned performance scores.
For example, given a different set of classification datasets where key class differences are the result of subtle variations in dynamical properties in long streams of time-series data, we would obtain a canonical set that might include features of long-range automutual information or measures the nonlinear time-series analysis literature: very different features to the relatively simple measures contained in \textit{catch22}.
As new time-series datasets are added to the UCR/UEA repository, that better capture the diversity of time-series data studied across industry and science, our feature reduction method could be rerun to extract new canonical feature sets that reflect the types of time-series properties that are important to measure in the new data.
Note that hybrid methods such as COTE \cite{Bagnall2016Time-seriesEnsembles}, which are not limited to a single time-series representation but can adapt to the problem at hand, consistently outperform both the shape-based existing classifiers and our features at the price of a much higher computational effort.
Given its computational efficiency, \textit{catch22} could be incorporated straightforwardly in these ensemble-based frameworks of multiple representations.
Here we excluded features that are sensitive to the location and spread of the data distribution, to ensure a fair comparison to shape-based methods which use normalized data; but for many real-world applications these could be highly relevant and should therefore be retained to allow improvements in classification accuracy.

In conclusion, here we present \textit{catch22}, a concise, accessible feature-based summary of an interdisciplinary time-series analysis literature for use in time-series classification tasks.
We hope that the ability to readily leverage feature-based representations of time series---and to generate new reduced feature sets tailored to specific domain problems---will aid diverse applications involving time series.


\section*{Information sharing statement}

The C-implementation of our canonical features along with their wrapped versions for R, Python and Matlab can be accessed on GitHub repository \\
\verb|https://github.com/chlubba/catch22|.\\

\noindent The selection pipeline is accessible on \\
\verb|https://github.com/chlubba/op_importance|.

\section*{Acknowledgements}

CL thanks EPSRC grant EP/L016737/1 and Galvani Bioelectronics.
SSS is supported by the Natural Environment Research Council through the Science and Solutions for a Changing Planet DTP.
BDF is supported by the NHMRC grant, 1089718.
NJ thanks EPSRC grants EP/N014529/1 and EP/K503733/1.

\newpage
\setcounter{table}{0}
\setcounter{figure}{0}
\setcounter{section}{0}
\setcounter{subsection}{0}
\renewcommand{\thesection}{S\arabic{section}}
\renewcommand{\thesubsection}{S\arabic{subsection}}
\renewcommand{\thetable}{S\arabic{table}}   
\renewcommand{\thefigure}{S\arabic{figure}}
\section*{Supplement}

\subsection{Insignificant features}

The features listed in Tab.~\ref{tab:insignificantops} were found to exhibit a classification performance across tasks consistent with a random-number generator.

{\scriptsize
\begin{longtable}{ll}
  \caption{
  The 145 features listed here exhibited classification performance consistent with a random-number generator.
  }
  \endlastfoot
  \toprule
  $p$-value & name\\ 
  \midrule 
  1.000 & WL\_coeffs\_db3\_4.wb99m \\
1.000 & WL\_coeffs\_db3\_3.wb99m \\
1.000 & WL\_coeffs\_db3\_2.wb99m \\
1.000 & WL\_coeffs\_db3\_1.wb99m \\
1.000 & Y\_LocalGlobal\_unicg500.iqr \\
1.000 & SP\_Summaries\_welch\_rect.linfitloglog\_hf\_sigrat \\
1.000 & SP\_Summaries\_pgram\_hamm.linfitloglog\_mf\_sigrat \\
1.000 & SP\_Summaries\_pgram\_hamm.linfitloglog\_hf\_sigrat \\
1.000 & SP\_Summaries\_pgram\_hamm.linfitloglog\_all\_sigrat \\
1.000 & SP\_Summaries\_fft\_logdev.logstd \\
1.000 & SP\_Summaries\_fft\_logdev.logiqr \\
1.000 & SP\_Summaries\_fft\_logdev.linfitsemilog\_all\_sigrat \\
1.000 & SP\_Summaries\_fft\_logdev.linfitsemilog\_all\_sigma \\
1.000 & SP\_Summaries\_fft\_logdev.linfitsemilog\_all\_sea1 \\
1.000 & SP\_Summaries\_fft\_logdev.linfitsemilog\_all\_a2 \\
1.000 & SP\_Summaries\_fft\_logdev.linfitloglog\_mf\_sigrat \\
1.000 & SP\_Summaries\_fft\_logdev.linfitloglog\_mf\_sigma \\
1.000 & SP\_Summaries\_fft\_logdev.linfitloglog\_mf\_sea1 \\
1.000 & SP\_Summaries\_fft\_logdev.linfitloglog\_mf\_a2 \\
1.000 & SP\_Summaries\_fft\_logdev.linfitloglog\_lf\_sigrat \\
1.000 & SP\_Summaries\_fft\_logdev.linfitloglog\_hf\_sigrat \\
1.000 & SP\_Summaries\_fft\_logdev.linfitloglog\_hf\_sigma \\
1.000 & SP\_Summaries\_fft\_logdev.linfitloglog\_hf\_sea1 \\
1.000 & SP\_Summaries\_fft\_logdev.linfitloglog\_hf\_a2 \\
1.000 & SP\_Summaries\_fft\_logdev.linfitloglog\_all\_sigrat \\
1.000 & SP\_Summaries\_fft\_logdev.linfitloglog\_all\_sigma \\
1.000 & SP\_Summaries\_fft\_logdev.linfitloglog\_all\_sea1 \\
1.000 & SD\_TSTL\_surrogates\_1\_100\_3\_tc3.stdsurr \\
1.000 & SD\_TSTL\_surrogates\_1\_100\_3\_tc3.normpatponmax \\
1.000 & SD\_TSTL\_surrogates\_1\_100\_3\_tc3.meansurr \\
1.000 & SD\_TSTL\_surrogates\_1\_100\_3\_tc3.kspminfromext \\
1.000 & SD\_TSTL\_surrogates\_1\_100\_3\_tc3.ksphereonmax \\
1.000 & SD\_TSTL\_surrogates\_1\_100\_3\_tc3.ksiqrsfrommode \\
1.000 & SD\_TSTL\_surrogates\_1\_100\_2\_trev.meansurr \\
1.000 & NL\_TSTL\_acp\_1\_001\_025\_10\_05.macpfdrop\_8 \\
1.000 & NL\_TSTL\_acp\_1\_001\_025\_10\_05.macpfdrop\_7 \\
1.000 & NL\_TSTL\_acp\_1\_001\_025\_10\_05.macpfdrop\_6 \\
1.000 & NL\_TSTL\_acp\_1\_001\_025\_10\_05.macpfdrop\_5 \\
1.000 & NL\_TSTL\_acp\_1\_001\_025\_10\_05.macpfdrop\_3 \\
1.000 & NL\_TSTL\_LargestLyap\_n1\_01\_001\_3\_1\_4.vse\_rmsres \\
1.000 & NL\_TSTL\_LargestLyap\_n1\_01\_001\_3\_1\_4.vse\_meanabsres \\
1.000 & NL\_MS\_nlpe\_fnn\_mi.maxonmean \\
1.000 & NL\_MS\_nlpe\_fnn\_mi.acmnd0 \\
1.000 & NL\_MS\_nlpe\_fnn\_mi.ac3n \\
1.000 & NL\_MS\_nlpe\_2\_mi.maxonmean \\
1.000 & MS\_shannon\_4\_1t10.stdent \\
1.000 & MS\_shannon\_4\_1t10.maxent \\
1.000 & MS\_shannon\_3\_1t10.stdent \\
1.000 & MS\_shannon\_3\_1t10.maxent \\
1.000 & MS\_shannon\_2\_1t10.maxent \\
1.000 & MF\_armax\_3\_1\_05\_1.lastimprovement \\
1.000 & MF\_armax\_3\_1\_05\_1.ac3n \\
1.000 & MF\_armax\_2\_2\_05\_1.lastimprovement \\
1.000 & MF\_StateSpace\_n4sid\_3\_05\_1.ac3n \\
1.000 & MF\_StateSpace\_n4sid\_3\_05\_1.ac2n \\
1.000 & MF\_GP\_LocalPrediction\_covSEiso\_covNoise\_5\_3\_10\_beforeafter.minstderr\_run \\
1.000 & MF\_GP\_LocalPrediction\_covSEiso\_covNoise\_5\_3\_10\_beforeafter.maxstderr\_run \\
1.000 & MF\_GP\_LocalPrediction\_covSEiso\_covNoise\_10\_3\_20\_randomgap.minstderr\_run \\
1.000 & MF\_GP\_LocalPrediction\_covSEiso\_covNoise\_10\_3\_20\_randomgap.maxstderr\_run \\
1.000 & MF\_GP\_LocalPrediction\_covSEiso\_covNoise\_10\_3\_20\_frombefore.minstderr\_run \\
1.000 & MF\_CompareTestSets\_y\_ar\_4\_rand\_25\_01\_1.ac1s\_iqr \\
1.000 & CO\_AddNoise\_ac\_kraskov1\_4.ami\_at\_15 \\
0.997 & MS\_shannon\_4\_1t10.meanent \\
0.994 & NL\_TSTL\_acp\_1\_001\_025\_10\_05.macpfdrop\_9 \\
0.979 & MF\_armax\_2\_2\_05\_1.ac3n \\
0.963 & SP\_Summaries\_fft\_logdev.linfitloglog\_lf\_sigma \\
0.834 & MS\_shannon\_2t10\_2.stdent \\
0.830 & TSTL\_delaytime\_01\_1.stdtau \\
0.830 & MF\_steps\_ahead\_ar\_best\_6.ac1\_1 \\
0.810 & SP\_Summaries\_welch\_rect.linfitloglog\_mf\_sigrat \\
0.810 & SP\_Summaries\_fft\_logdev.linfitloglog\_lf\_sea1 \\
0.781 & NL\_TSTL\_acp\_1\_001\_025\_10\_05.macpfdrop\_1 \\
0.712 & SP\_Summaries\_welch\_rect.linfitloglog\_hf\_sigma \\
0.703 & SP\_Summaries\_welch\_rect.linfitloglog\_lf\_sigrat \\
0.703 & MF\_StateSpaceCompOrder\_8.maxdiffaic \\
0.699 & NL\_TSTL\_acp\_1\_001\_025\_10\_05.macpfdrop\_2 \\
0.699 & NL\_MS\_nlpe\_fnn\_mi.p3\_5 \\
0.697 & PH\_ForcePotential\_dblwell\_3\_001\_01.finaldev \\
0.608 & NL\_MS\_nlpe\_fnn\_mi.ac2n \\
0.561 & NL\_MS\_nlpe\_fnn\_mi.meane \\
0.561 & NL\_MS\_nlpe\_2\_mi.ac3n \\
0.511 & CO\_AddNoise\_ac\_kraskov1\_4.ami\_at\_10 \\
0.484 & CO\_AddNoise\_ac\_kraskov1\_4.ami\_at\_20 \\
0.432 & WL\_coeffs\_db3\_max.wb99m \\
0.375 & NL\_MS\_nlpe\_fnn\_mi.p5\_5 \\
0.354 & SP\_Summaries\_fft\_logdev.linfitloglog\_hf\_a1 \\
0.354 & NL\_MS\_nlpe\_fnn\_mi.p2\_5 \\
0.300 & NL\_MS\_nlpe\_2\_mi.p4\_5 \\
0.295 & SP\_Summaries\_welch\_rect.linfitloglog\_hf\_sea1 \\
0.286 & NL\_TSTL\_acp\_1\_001\_025\_10\_05.stdmacpfdiff \\
0.286 & CO\_StickAngles\_y.ratmean\_p \\
0.278 & CO\_AddNoise\_ac\_std1\_10.ami\_at\_10 \\
0.261 & MF\_armax\_2\_2\_05\_1.ac2n \\
0.227 & NL\_MS\_nlpe\_2\_mi.acmnd0 \\
0.223 & NL\_MS\_nlpe\_fnn\_mi.ac3 \\
0.222 & MF\_CompareTestSets\_y\_ar\_best\_uniform\_25\_01\_1.ac1s\_iqr \\
0.222 & CO\_AddNoise\_1\_std1\_10.pdec \\
0.218 & CO\_AddNoise\_1\_kraskov1\_4.ami\_at\_15 \\
0.183 & CO\_AddNoise\_ac\_std1\_10.ami\_at\_15 \\
0.172 & Y\_LocalGlobal\_unicg100.iqr \\
0.170 & MF\_armax\_2\_2\_05\_1.p5\_5 \\
0.166 & SP\_Summaries\_fft\_logdev.linfitloglog\_all\_a2 \\
0.152 & Y\_SlidingWindow\_lil\_ent10\_1 \\
0.145 & PH\_ForcePotential\_dblwell\_2\_005\_02.finaldev \\
0.140 & MF\_StateSpace\_n4sid\_2\_05\_1.ac3n \\
0.133 & Y\_LocalGlobal\_unicg50.iqr \\
0.129 & NL\_MS\_nlpe\_fnn\_mi.acsnd0 \\
0.129 & CO\_AddNoise\_ac\_std1\_10.pdec \\
0.127 & CO\_AddNoise\_1\_kraskov1\_4.ami\_at\_20 \\
0.127 & MS\_shannon\_2t10\_2.medent \\
0.124 & MF\_CompareTestSets\_y\_ss\_best\_uniform\_25\_01\_1.ac1s\_iqr \\
0.119 & SP\_Summaries\_welch\_rect.linfitloglog\_all\_sigrat \\
0.105 & NL\_MS\_nlpe\_2\_mi.ac2n \\
0.104 & NL\_TSTL\_acp\_1\_001\_025\_10\_05.macpfdrop\_4 \\
0.096 & MS\_shannon\_2t10\_3.stdent \\
0.090 & MS\_shannon\_3\_2 \\
0.088 & MS\_shannon\_3\_1t10.meanent \\
0.083 & MF\_GP\_LocalPrediction\_covSEiso\_covNoise\_10\_3\_20\_frombefore.maxerrbar \\
0.080 & SP\_Summaries\_fft\_logdev.logarea\_3\_3 \\
0.080 & FC\_LoopLocalSimple\_mean.sws\_stdn \\
0.077 & SP\_Summaries\_fft\_logdev.logarea\_5\_5 \\
0.074 & MS\_shannon\_2t10\_4.stdent \\
0.074 & MF\_armax\_3\_1\_05\_1.ac2n \\
0.069 & SP\_Summaries\_fft\_logdev.logarea\_4\_4 \\
0.068 & ST\_LocalExtrema\_n25.minabsmin \\
0.068 & NL\_MS\_nlpe\_2\_mi.p3\_5 \\
0.068 & MS\_shannon\_4\_1t10.medent \\
0.068 & MF\_FitSubsegments\_arma\_2\_2\_uniform\_25\_01.q\_2\_std \\
0.068 & SP\_Summaries\_fft\_logdev.q25 \\
0.061 & NL\_MS\_nlpe\_2\_mi.p2\_5 \\
0.060 & MF\_FitSubsegments\_arma\_2\_2\_uniform\_25\_01.q\_1\_max \\
0.060 & MS\_shannon\_2\_1t10.stdent \\
0.059 & NL\_TSTL\_acp\_mi\_1\_\_10.iqracpf\_1 \\
0.059 & NL\_MS\_nlpe\_fnn\_mi.ac1n \\
0.059 & MF\_FitSubsegments\_arma\_2\_2\_uniform\_25\_01.q\_2\_min \\
0.059 & MF\_CompareTestSets\_y\_ss\_2\_uniform\_25\_01\_1.ac1s\_iqr \\
0.059 & NL\_MS\_nlpe\_2\_mi.p5\_5 \\
0.056 & MF\_StateSpace\_n4sid\_2\_05\_1.ac2n \\
0.056 & DN\_SimpleFit\_gauss2\_hsqrt.resAC2 \\
0.056 & CO\_AddNoise\_ac\_std1\_10.ami\_at\_20 \\
0.056 & SP\_Summaries\_fft\_logdev.logarea\_2\_2 \\
0.055 & MF\_FitSubsegments\_arma\_2\_2\_uniform\_25\_01.q\_2\_max \\
0.054 & NL\_TSTL\_acp\_mi\_1\_\_10.macpfdrop\_4 \\
0.050 & Y\_LocalGlobal\_unicg20.iqr \\
0.050 & NL\_MS\_nlpe\_2\_mi.ac2 \\
\bottomrule
\label{tab:insignificantops}
\end{longtable}
}

\clearpage
\subsection{Time series for computation time evaluation}
\label{sec:empiricalts}

A selection of 40 time series was obtained from the dataset `1000 Empirical Time series' \cite{Fulcher20171000Series}.

\begin{table}[H]
\scriptsize
        \begin{tabular}[t]{lll} 
        \toprule
        ID & name & keywords\\ 
        \midrule 
1 & NS\_beta\_L10000\_a1\_b3\_2.dat & synthetic,noise,beta \\ 
25 & ST\_M5a\_N10000\_a-0.01\_b-0.6\_c0\_d0\_x0\_1\_2.dat & synthetic,stochastic,SDE,M5 \\ 
53 & SY\_rwalk\_L10000\_20.dat & synthetic,randomwalk \\ 
75 & FL\_ddp\_L300\_N5000\_IC\_0.1\_0.1\_y.dat & synthetic,dynsys,ddp \\ 
106 & FL\_lorenz\_L250\_N10000\_IC\_0\_-0.01\_9.1\_y.dat & synthetic,dynsys,chaos,lorenz \\ 
125 & FL\_shawvdp\_L300\_N5000\_IC\_1.3\_0.1\_x.dat & synthetic,dynsys,shawvdp \\ 
158 & MP\_burgers\_L300\_IC\_-0.2\_0.1\_x.dat & synthetic,map,burgers \\ 
175 & MP\_chirikov\_L300\_IC\_0.2\_6\_y.dat & synthetic,map,chirikov \\ 
211 & MP\_henon\_L1000\_a1.4\_b0.3\_IC\_0\_0.9.dat & synthetic,map,henon \\ 
225 & MP\_holmescubic\_L300\_IC\_1.7\_0\_y.dat & synthetic,map,holmescubic \\ 
263 & MP\_lorenz3d\_L300\_IC\_0.51\_0.5\_-1\_x.dat & synthetic,map,chaos,lorenz3d \\ 
275 & MP\_pinchers\_L5000\_s2.1c\_0.55.dat & synthetic,map,pinchers \\ 
316 & MP\_spence\_L5000\_x0\_0.27.dat & synthetic,map,spence \\ 
325 & MP\_tent\_L5000\_A1.88.dat & synthetic,map,tent,chaos \\ 
369 & SY\_MA\_L500\_p8\_7.dat & synthetic,MA,MA8 \\ 
375 & SY\_MIX\_p0.3\_L5000\_5.dat & synthetic,MIXP,MIX0.3 \\ 
421 & FI\_yahoo\_HL\_KLSE.dat & finance,yahoo,opening \\ 
425 & FI\_yahoo\_HL\_z28\_SPMIB.dat & finance,yahoo,opening \\ 
474 & FL\_dblscroll\_L1000\_N5000\_IC\_0.01\_0.01\_0\_z.dat & synthetic,dynsys,chaos,dblscroll \\ 
475 & FL\_dblscroll\_L200\_N10000\_IC\_0.01\_0.01\_0\_z.dat & synthetic,dynsys,chaos,dblscroll \\ 
525 & FL\_moorespiegel\_L250\_N1000\_IC\_0.1\_0\_0\_x.dat & synthetic,dynsys, \\ 
& & chaos,moorespiegel \\ 
526 & FL\_moorespiegel\_L250\_N5000\_IC\_0.1\_0\_0\_x.dat & synthetic,dynsys, \\ 
& & chaos,moorespiegel \\ 
575 & FL\_simpqcf\_L1000\_N10000\_IC\_-0.9\_0\_0.5\_z.dat & synthetic,dynsys,chaos,simpqcf \\ 
579 & FL\_simpqcf\_L2000\_N1000\_IC\_-0.9\_0\_0.5\_z.dat & synthetic,dynsys,chaos,simpqcf \\ 
625 & MP\_Lozi\_iii.dat & synthetic,map,chaos,lozi \\ 
631 & MP\_freitas\_nlma\_L500\_a-3.92\_b-3.1526\_1.dat & synthetic,map,nonlinear,freita \\ 
675 & TR\_arge030\_rf.dat & treerings \\ 
684 & AS\_s2.2\_f4\_b8\_l8800\_58925.dat & sound,animalsounds \\ 
725 & FI\_yahoo\_Op\_IFL.L\_LOGR.dat & finance,logr \\ 
737 & Ich\_recflow.dat & meteorology,riverflow, \\ 
& & reconstructed,UK \\ 
775 & SF\_E\_5.dat & SantaFe,astronomy \\ 
789 & SPIDR\_hpidmsp\_F15\_meas.dat & space,hpidmsp \\ 
825 & SPIDR\_meanDelay\_ACE\_hrly.dat & space,magneticfield \\ 
842 & SPIDR\_vostok\_L6600\_Sep2002\_vostok.dat & space,vostok \\ 
875 & t\_osaka\_rf.dat & meteorology,temperature \\ 
894 & MD\_chfdb\_chf07\_seg039\_SNIP\_9574-17073.dat & medical,physionet,ecg,chfdb,snip \\ 
925 & MD\_tremordb\_g12ren.dat & medical,tremor,physionet,lowamp, \\ 
& & dbson,medon,gpi \\ 
947 & MUS\_Tetra-Sync\_1364s\_F0.02\_b8.dat & sound,music,downed \\ 
976 & MD\_nsrdb\_nsr19088\_seg007\_SNIP\_5659-15658.dat & medical,physionet,ecg,nsrdb,snip \\ 
1000 & MD\_mghdb\_mgh79\_PAP\_SNIP\_9047-15146.dat & medical,physionet,mghdb, \\ 
& & snip,pulmonaryarterialpressure \\ 
        \bottomrule
        \end{tabular}
        \captionof{table}{40 empirical time series selected for evaluating the computation times of features.}
        \label{tab:empiricalts}
\end{table}

\clearpage
\subsection{Performance comparison with \textit{tsfeatures}}

\begin{figure}[H]
\centering
  \includegraphics[width=7cm]{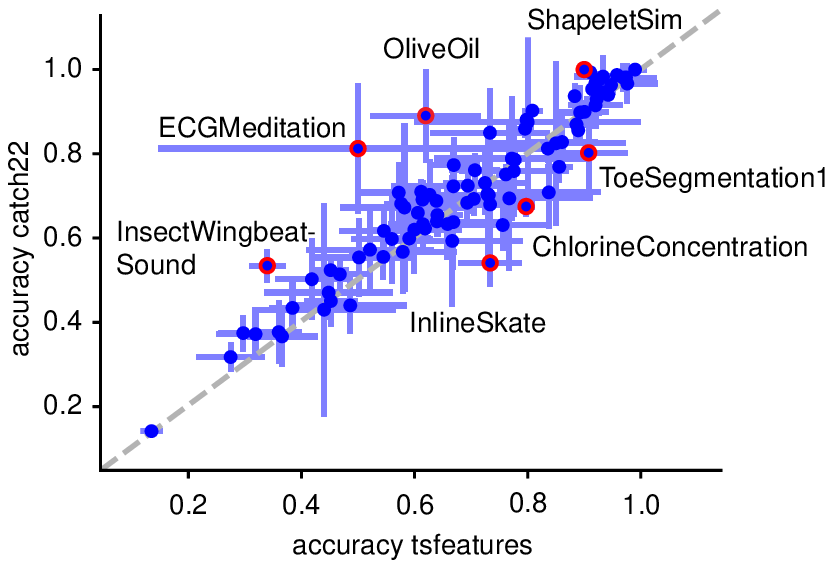}
\caption{
\textbf{Our automatically selected \textit{catch22} feature set performs as well as the standard feature set for simple time series contained in the \textit{tsfeatures} package.} Class-balanced accuracy is shown for \textit{tsfeatures} and \textit{catch22}, error bars indicate standard deviation across folds.
A gray dashed equality line is annotated, and particular datasets with the greatest differences in accuracy are highlighted as red circles and labeled.
}
\label{fig:hyndman}
\end{figure}

\bibliographystyle{spmpsci.bst}

\end{document}